\documentclass[multphys,vecphys]{svmult}
 
\usepackage{makeidx}         % allows index generation
\usepackage{graphicx}        % standard LaTeX graphics tool
\usepackage{exscale,relsize}
\usepackage{epsf,color,graphics}
\usepackage{colordvi}
\usepackage{multicol}        % used for the two-columsit index
\usepackage[bottom]{footmisc}% places footnotes at page bottom
\usepackage{amsmath}
\usepackage{amssymb}
\usepackage{txfonts,captcont,subfigure}
\usepackage{rotating}
\usepackage{natbib}      % bibliography at the end
\usepackage{wrapfig}
 
\usepackage{natbib} % bibliography
\usepackage[below]{placeins}
\usepackage{url}
 
\def\etal{{\it et al}.}

\def\d3k{{\displaystyle {\rm d}{\bmit k} \over \displaystyle (2\pi)^3}}
\def\d3x{{d{\bf x}}}

\def\d3k{{\displaystyle {\rm d}{\bf k} \over \displaystyle (2\pi)^3}}

\def\spose#1{\hbox to 0pt{#1\hss}}
\def\lta{\mathrel{\spose{\lower 3pt\hbox{$\mathchar"218$}}
     \raise 2.0pt\hbox{$\mathchar"13C$}}}
\def\gta{\mathrel{\spose{\lower 3pt\hbox{$\mathchar"218$}}
     \raise 2.0pt\hbox{$\mathchar"13E$}}}
\font \bigtwo=cmbx10 scaled\magstep2

\makeindex             % used for the subject index
\begin{document}
 
\title*{\bigtwo Cosmic Order out of Primordial Chaos: \\ a tribute to Nikos Voglis}
% Use 
\titlerunning{Cosmic Order out of Primordial Chaos} 
%for an abbreviated version of
% your contribution title if the original one is too long
\author{Bernard Jones \& Rien van de Weygaert}
% Use \authorrunning{Short Title} for an abbreviated version of
% your contribution title if the original one is too long
\institute{Kapteyn Astronomical Institute, University of Groningen, P.O. Box 800, 
  9700 AV Groningen, the Netherlands
\ \\
\texttt{bernard@astrag.demon.co.uk,weygaert@astro.rug.nl}}
%
% Use the package "url.sty" to avoid
% problems with special characters
% used in your e-mail or web address
%
\maketitle
 
\begin{abstract}
Nikos Voglis had many astronomical interests, among them was the question of the origin of galactic angular momentum.  In this short tribute we review how this subject has changed since the 1970's and how it has now become evident that gravitational tidal forces have not only caused galaxies to rotate, but have also acted to shape the very cosmic structure in which those galaxies are found.  We present recent evidence for this based on data analysis techniques that provide objective catalogues of clusters, filaments and voids.
\end{abstract}
 
%============================
\section{Some early history}
%============================
%\begin{figure*}[t]
\begin{wrapfigure}[21]{r}{1.5in}
     \vskip -0.25in
     \includegraphics[width=1.5in]{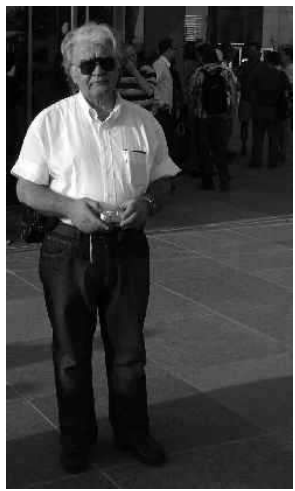}
	 \caption{Nikos at the ``bernard60'' conference (Valencia, June 2006). Picture taken by Phil Palmer.}
\end{wrapfigure}  
It was in the 1970's that Nikos Voglis first came to visit Cambridge, England, to attend a conference and to discuss a problem that was to remain a key area of personal interest for many years to come: the origin of galaxy angular momentum.  It was during this period that Nikos teamed up with Phil Palmer to create a long lasting and productive collaboration.
 
The fundamental notion that angular momentum is conserved leads one to wonder how galaxies could acquire their angular momentum if they started out with none. This puzzle was perhaps one of the main driving forces behind the idea that cosmic structure was born out of some primordial turbulence.  However, by the early 1970's the cosmic turbulence theory was falling into disfavour owing to a number of inherent problems (see \citet{Jones76} for a detailed review of this issue).  
 
The alternative, and now well entrenched, theory was the gravitational instability theory in which structure grew through the driving force of gravitation acting on primordial density perturbations.  The question of the origin of angular momentum had to be addressed and would be central to the success or failure of that theory.  \citep{Peebles69} provided the seminal paper on this, proposing that tidal torques would be adequate to provide the solution.  However, this was for many years mired in controversy.
 
Tidal torques had been suggested as a source for the origin of angular momentum since the late 1940's when \citet{Hoyle49} invoked the tidal stresses exerted by a cluster on a galaxy as the driving force of galaxy rotation.  Although the idea as expounded was not specific to any cosmology, there can be little doubt that Hoyle had his Steady State cosmology in mind. The \citet{Peebles69} version of this process specifically invoked the tidal stresses between two neighbouring protogalaxies, but it was not without controversy. 
There were perhaps three sources for the ensuing debate:
\begin{itemize}
\item Is the tidal force sufficient to generate the required angular momentum>
\item Are tidal torques between proto-galaxies alone responsible for the origin of galactic angular momentum?
\item Tidal torques produce shear fields – what is the origin of the observed circular rotation?
\end{itemize}
\cite{oort1970} and \cite{harrison1971} had both argued that the interaction between low-amplitude primordial perturbations would be inadequate to drive the rotation: they saw the positive density fluctuations as being ``shielded'' by a surrounding negative density region which would diminish the tidal forces.  This doubt was a major driving force behind ``alternative'' scenarios for galaxy formation.  The last of these was a more subtle problem since, to some, even if tidal forces managed to generate adequate shear flows, the production of rotational motion would nonetheless require some violation of the Kelvin circulation theorem.  Although the situation was clarified by \citet{Jones76} it was not until the exploitation of N-Body cosmological simulations that the issue was considered to have been resolved. 
 
It was into this controversy that Nikos stepped, asking precisely these questions.  A considerable body of his later work (much of it with Phil Palmer, see for example \citet{PalVog83}) was devoted to addressing these issues at various levels. Since these days our understanding of the tidal generation of 
galaxy rotation has expanded impressively, mostly as a result of ever more sophisticated and large N-body simulation \citep[e.g.][]{efstathjones79,
jonesefstath79,Barnes87,Porciani02,Bosch02,Bett07}. What remains is Nikos' urge for a deeper insight, beyond simulation, into the physical intricacies 
of the problem. 
  
%=========================================================
\section{Angular Momentum generation: the tidal mechanism}
%=========================================================
\label{sec:tidaltorque}
\begin{figure*}
  \vskip -0.25truecm
  \begin{center}
  \includegraphics[width=3.6in]{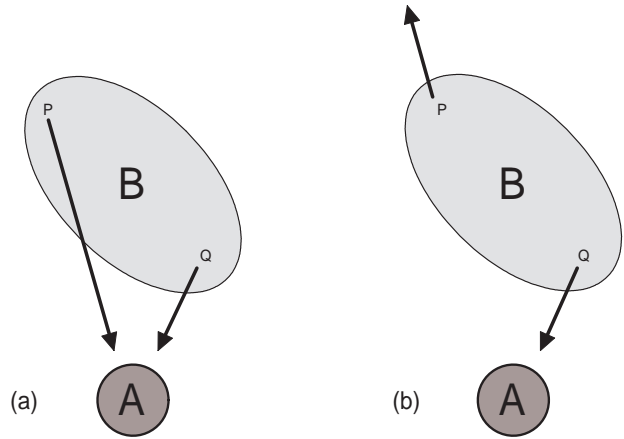}  
  \caption{An extended object, {\bf B}, acted on by the gravitational field of a nearby object, {\bf A}.  (a) depicts the forces as seen from the point of view of the forcing object: both points P and Q fall towards the A, albeit at different rates.  (b) depicts the forces as seen from the point of view of the mass center of {\bf B} where both P and Q recede from the mass center.}
\vskip -0.75truecm
  \end{center}
  \label{fig:torques}
\end{figure*}
In order to appreciate these problems it is helpful to look at a simplified version of the tidal model as proposed by Peebles.  Consider two neighbouring, similar sized, protogalaxies {\bf A} and {\bf B} (figure ~\ref{fig:torques}).  We can view the tidal forces exerted on {\bf B} by {\bf A} from either the reference frame of the mass center of {\bf A} or from the reference frame of the mass center of {\bf B} itself.  These forces are depicted by arrows in the figure: note that relative to the mass center of {\bf B} the tidal forces act so as to stretch {\bf B} out in the direction of {\bf A}.
 
To a first approximation, the force gradient acting on {\bf B} can be expressed in terms of the potential field $\phi(\bf x)$ in which {\bf B} is situated:
\begin{equation}
T_{ij} = \displaystyle {\partial {F_i} \over \partial {x_j}} = \displaystyle {{{\partial^2 \phi} \over {\partial x_i \partial x_j}}} - {1 \over 3} \delta_{ij} \nabla^2 \phi
\label{eq:tidetensor}
\end{equation}
where the potential field is determined from the fluctuating component of the density field via the Poisson equation \footnote{The Poisson equation determines only the trace of the symmetric tensor $\displaystyle {{{\partial^2 \phi} \over {\partial x_\alpha \partial x_\beta}}}$.  The interesting exercise for the reader is to contemplate what determines the other 5 components?}.  The flow of material is thus a shear flow determined by the principal directions and magnitudes of inertia tensor of the blob {\bf B}.  Viewed as a fluid flow this is undeniably a shear flow with zero vorticity as demanded by the Kelvin circulation theorem \footnote{This raises the technically interesting question as to whether a body with zero angular momentum can rotate: most undergraduates following a classical dynamics course with a section on rigid bodies would unequivocally answer ``no''.  The situation is beautifully discussed in Feynman's famous ``Lectures in Modern Physics'' \citep{Feynman70}.}.  
 
So how does the vorticity that is evident in galaxy rotation arise?  The answer is twofold.  Shocks will develop in the gas flow and stars will form: the Kelvin Theorem holds only for nondissipative flows.  Then, a ``gas'' of stars does not obey the Kelvin Theorem since it is not a fluid (though there is a six-dimensional phase space analogue for a stellar ``gas'').
 
The magnitude and direction of angular momentum vector is related to the inertia tensor. $I_{ij}$, of the torqued object and the driving tidal forces described by the tensor $T_{ij}$ of equation (\ref{eq:tidetensor}).  In 1984, based on simple low-order perturbation theory, \citet{White84} wrote an intuitively appealing expression for the angular momentum vector $L_i$ of a protogalaxy having inertia tensor $I_{mk}$:
\begin{equation}
L_i \propto \epsilon_{ijk} T_{jm} I_{mk},
\label{eq:angmom}
\end{equation}
where summation is implied over the repeated indices.  This was later taken up by \citet{CatTheuns99} in a high-order perturbation theory discussion of the problem.  However, there is in these treatments an underlying assumption, discussed but dismissed by \citet{CatTheuns99}, that the tensors $T_{ij}$ and $I_{ij}$ are statistically independent. Subsequent numerical work by \citet{LeePen00} showed that this assumption is not correct and that ignoring it results in an incorrect estimator for the magnitude of the spin.
 
The approach taken by \citet{LeePen00, LeePen01} is interesting: they write down an equation for the autocorrelation tensor of the angular momentum vector in a given tidal field, averaging over all orientations and magnitudes of the inertia  tensor. On the basis of equation (\ref{eq:angmom}) one would expect this tensor autocorrelation function to be given by
\begin{equation} 
\langle L_i L_j | {\bf T} \rangle \propto \epsilon_{ipq} \epsilon_{jrs}  T_{pm}T_{rn} \langle I_{mq} I_{ns}\rangle
\end{equation}
where the notation $\langle L_i L_j | {\bf T} \rangle$ is used to emphasise that $T_{ij}$ is regarded as a given value and is not a random variable.  The argument then goes that the isotropy of underlying density distribution allows us to replace the statistical quantity $\langle I_{mq} I_{ns}\rangle$ by a sum of Kronecker deltas leaving only
\begin{equation}
\langle L_i L_j | {\bf T} \rangle \propto {1 \over 3} \delta_{ij} + ({1 \over 3} \delta_{ij} - T_{ik}T_{kj} )
\label{eq:indepIT}
\end{equation}
It is then asserted that if the moment of inertia and tidal shear tensors were uncorrelated, we would have only the first term on the right hand side, 
${1 \over 3} \delta_{ij}$: the angular momentum vector would be isotropically distributed relative to the tidal tensor. 

In fact, in the primordial density field and the early linear phase of structure formation there is a significant 
correlation between the shape of density fluctuations and the tidal force field \citep{bond1987,weyedb1996}. Part of the 
correlation is due to the anisotropic shape of density peaks and the internal tidal gravitational force field that goes 
along with it \citep{icke1973}. The most significant factor is that of intrinsic spatial correlations in the primordial 
density field. It is these intrinsic correlations between shape and tidal field that are at the heart of our understanding 
of the Cosmic Web, as has been recognized by the Cosmic Web theory of \cite{bondweb96}. The subsequent nonlinear evolution 
may strongly augment these correlations (see e.g. fig.~\ref{fig:cosmicwebtide}), although small-scale highly nonlinear 
interactions also lead to a substantial loss of the alignments: clusters are still strongly aligned, while galaxies seem 
less so. 

Recognizing that the inertia and tidal tensors may not be mutually independent, \cite{LeePen00,LeePen01} write
\begin{equation}
\langle L_i L_j | {\bf T} \rangle \propto {1 \over 3} \delta_{ij} + c ({1 \over 3} \delta_{ij} - T_{ik}T_{kj} )
\end{equation}
where $c = 0$ for randomly distributed angular momentum vectors.  The case of mutually independent tidal and inertia tensors is described by $c = 1$ (see equation \ref{eq:indepIT}).  They finally introduce a different parameter $a = 3c/5$ and write
\begin{equation}
\langle L_i L_j | {\bf T} \rangle \propto {{1+a} \over 3} \delta_{ij} - a T_{ik}T_{kj} 
\end{equation}
which forms the basis of much current research in this field.  The value derived from recent study of the Millenium simulations by \citet{LeePen07} is $a \approx 0.1$. 
  
%============================
\section{Gravitational Instability}
%============================
In the gravitational instability scenario, \citep[e.g.][]{peebles1980}, cosmic structure grows from an intial random field of primordial density and velocity perturbations. The formation and molding of structure is fully described by three equations, the {\it continuity equation}, expressing mass conservation, the {\it Euler equation} for accelerations driven by the gravitational force for dark matter and gas, and pressure forces for the gas, and the {\it Poisson-Newton equation} relating the gravitational potential to the density.
 
A general density fluctuation field for a component of the universe with respect to its cosmic background mass density $\rho_{\rm u}$ is defined by
\begin{equation}
\delta({\bf r},t)\,=\,\frac{\rho({\bf r})-\rho_{\rm u}}{\rho_{\rm u}}\,.
\end{equation}
Here ${\bf r}$ is comoving position, with the average expansion factor $a(t)$ of the universe taken out. Although there are fluctuations in photons, neutrinos, dark energy, etc., we focus here on only those contributions to the mass which can cluster once the relativistic particle contribution has become small, valid for redshifts below 100 or so. A non-zero $\delta({\bf r},t)$ generates a corresponding total peculiar gravitational acceleration ${\bf g}({\bf r})$ which at any cosmic position ${\bf r}$ can be written as the integrated effect of the peculiar gravitational attraction exerted by all matter
fluctuations throughout the Universe:
\begin{equation}
{\bf g}({\bf r},t)\,=\,- 4\pi G \bar{\rho}_m
(t)a(t) \int {\rm d}{\bf r}'\,\delta({\bf r}^\prime,t)\,{\displaystyle
({\bf r}-{\bf r}^\prime) \over \displaystyle |{\bf r}-{\bf
r}^\prime|^3}\ .
\label{eq:gravstab} 
\end{equation}
Here $\bar{\rho}_{m}(t)$ is the mean density of the mass in the universe that can cluster (dark matter and baryons). The
cosmological density parameter $\Omega_m (t)$ is defined by $\rho_{\rm u}$, via the relation $\Omega_m H^2 = (8\pi G/3)
\bar{\rho_{m}} $ in terms of the Hubble parameter $H$. The relation between the density field and gravitational potential $\Phi$ is established through the Poisson-Newton equation:
\begin{equation}
\nabla^2 \Phi = 4 \pi G \bar{\rho}_{m}(t) a(t)^2 \ \delta({\bf r},t) . 
\end{equation}
The peculiar gravitational acceleration is related to $\Phi({\bf r},t)$ through ${\bf g}=-\nabla \Phi/a$ and drives peculiar motions.
\begin{figure*}[t]
  \vskip -0.2truecm
  \begin{center}
     \mbox{\hskip -0.5truecm\includegraphics[width=12.5cm]{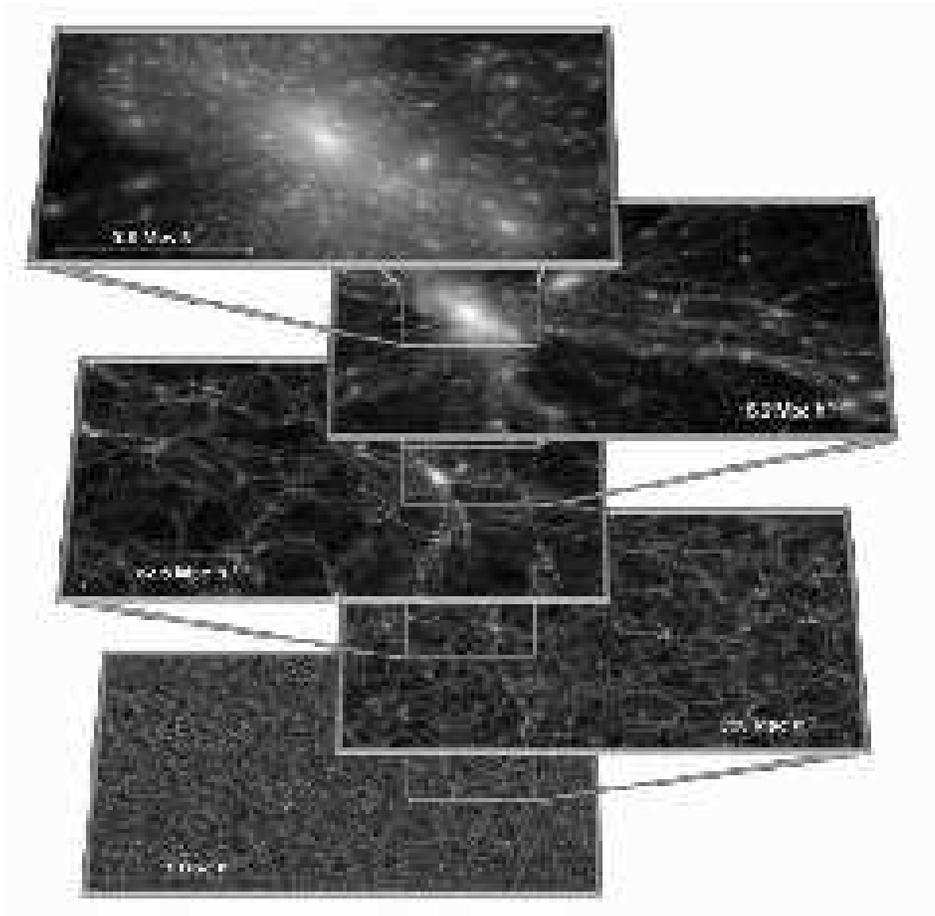}}
    \vskip 0.25cm
  \caption{The hierarchical Cosmic Web: over a wide range of spatial and mass scales structures and features are embedded within structures of a larger effective dimension and a lower density.  Image courtesy of V. Springel \& Virgo consortium, also
see Springel et al. 2005. Reproduced with permission of Nature.}
  \end{center}
\label{fig:millenniumhier}
\vskip -0.75truecm
\end{figure*}
In slightly overdense regions around density excesses, the excess gravitational attraction slows down the expansion relative to the mean, while underdense regions expand more rapidly. The underdense regions around density minima expand relative to the background, forming deep voids. Once the gravitational clustering process has progressed beyond the initial linear growth phase we see the emergence of complex patterns and structures in the density field. 

Large N-body simulations all reveal a few ``universal'' characteristics of the (mildly) nonlinear cosmic matter distribution: its hierarchical nature, the anisotropic and weblike spatial geometry of the spatial mass distribution and the presence of huge underdense voids.  These basic elements of the Cosmic Web \citep{bondweb96,weybondgh2008} exist at all redshifts, but differ in scale.  

Fig.~\ref{fig:millenniumhier}, from the state-of-the-art ``Millennium simulation'', illustrates this complexity in great detail over a substantial range of scales.  The figure zooms in on the dark matter distribution at five levels of spatial resolution and shows the formation of a filamentary network connecting to a central cluster.  This network establishes transport channels along which matter will flow into the cluster.  The hierarchical nature of the structure is clearly visible. The dark matter distribution is far from homogeneous: a myriad of tiny dense clumps indicate the presence of dark halos in which galaxies, or groups of galaxies, will have formed. 

Within the context of gravitational instability, it is the gravitational tidal forces that establish the relationship between some of the most prominent manifestations of the structure formation process. It is this intimate link between the Cosmic Web, 
the mutual alignment between cosmic structures and the rotation of galaxies to which we wish to draw attention in this short contribution. 

\section{Tidal Shear}
When describing the dynamical evolution of a region in the density field it is useful to distinguish between large scale ``background'' fluctuations $\delta_{\rm b}$ and small-scale fluctuations $\delta_{\rm f}$. Here, we are primarily interested in the influence of the smooth large-scale field. Its scale $R_b$ should be chosen such that it remains (largely) linear, i.e. the r.m.s. density fluctuation amplitude $\sigma_{\rho}(R_b,t) \lesssim 1$. 

To a good approximation the smoother background gravitational force ${\bf g}_{\rm b}({\bf x})$ (eq.~\ref{eq:gravstab}) in and around the mass element includes three components (apart from rotational aspects). The {\it bulk force} ${\bf g}_{\rm b}({\bf x}_{pk})$ is responsible for the acceleration of the mass element as a whole. Its divergence ($\nabla \cdot {\bf g}_{\rm b}$) encapsulates the collapse of the overdensity while the tidal tensor $T_{ij}$ quantifies its deformation,  
\begin{equation}
g_{\rm b,i}({\bf x})\,=\,g_{\rm b,i}({\bf x}_{pk})\,+\,a\,\sum_{j=1}^3\,\left\{{\displaystyle 1 \over \displaystyle 3 a}(\nabla \cdot {\bf g}_{\rm b})({\bf x}_{pk})\,\delta_{\rm ij}\,-\,T_{ij}\right\} (x_j - x_{pk, {\rm j}})\,.
\end{equation}
The tidal shear force acting over the mass element is represented by the (traceless) tidal tensor $T_{ij}$,
\begin{eqnarray}
T_{\rm ij}&\,\equiv\,&-{\displaystyle 1 \over \displaystyle 2 a}\, 
\left\{ {\displaystyle \partial g_{{\rm b},i} \over \displaystyle \partial x_i} +
{\displaystyle \partial g_{{\rm b},j} \over \displaystyle \partial x_j}\right\}\,+\,
{\displaystyle 1 \over \displaystyle 3 a} (\nabla \cdot {\bf g}_{\rm b}) 
\,\delta_{\rm ij}
\end{eqnarray}
in which the trace of the collapsing mass element, proportional to its overdensity $\delta$, dictates its contraction (or expansion). For a cosmological matter distribution the close connection between local force field and global matter distribution follows from the expression of the tidal tensor in terms of the generating cosmic matter density fluctuation distribution $\delta({\bf r})$ 
\citep{weyedb1996}:
\begin{eqnarray}
&& T_{ij}({\bf r})={\displaystyle 3 \Omega H^2 \over \displaystyle 8\pi}\,
\int {\rm d}{\bf r}'\,\delta({\bf r}')\ \left\{{\displaystyle 3 (r_i'-r_i)(r_j'-r_j) -
|{\bf r}'-{\bf r}|^2\ \delta_{ij} \over \displaystyle |{\bf r}'-{\bf r}|^5}\right\} - {\frac{1}{2}}\Omega H^2\ \delta({\bf r},t)\ \delta_{ij} . \nonumber 
\label{eq:quadtide}
\end{eqnarray}
The tidal shear tensor has been the source of intense study by the gravitational lensing community since it is now possible to 
map the distribution of large scale cosmic shear using weak lensing data. See for example \citet{HirSel04,massey2007}.
\begin{figure*}[t]
\begin{center}
\vskip 0.25truecm
\mbox{\hskip -2.5truecm\includegraphics[width=12.1cm]{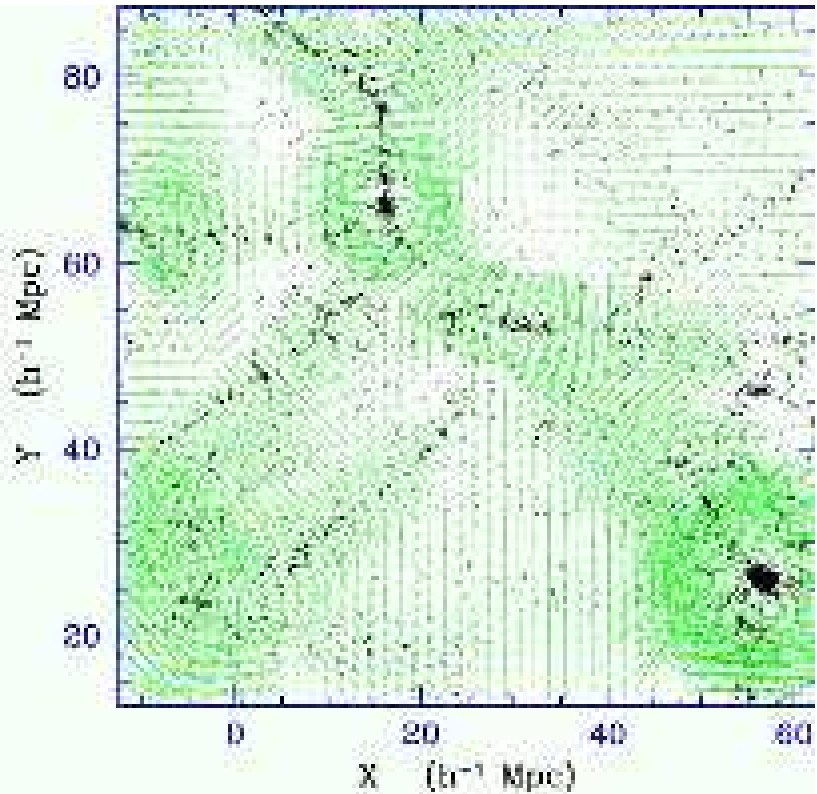}}
\vskip -5.0truecm
\caption{The relation between the {\it cosmic web}, the clusters at the nodes in this network and 
the corresponding compressional tidal field pattern. It shows the matter distribution at the 
present cosmic epoch, along with the (compressional component) tidal field bars in a slice through 
a simulation box containing a realization of cosmic structure formed in an open, $\Omega_{\circ}=0.3$, Universe 
for a CDM structure formation scenario (scale: $R_G=2h^{-1}\hbox{\rm Mpc}$). The frame shows structure in a 
$5h^{-1}\hbox{\rm Mpc}$ thin central slice, on which the related tidal bar configuration is superimposed. 
The matter distribution, displaying a pronounced weblike geometry, is clearly intimately linked with a 
characteristic coherent compressional tidal bar pattern. From: van de Weygaert 2002} 
\label{fig:cosmicwebtide}
\end{center}
\end{figure*}
\section{The Cosmic Web}
Perhaps the most prominent manifestation of the tidal shear forces is that of the distinct {\it weblike 
geometry} of the cosmic matter distribution, marked by highly elongated filamentary, flattened planar structures 
and dense compact clusters surrounding large near-empty void regions (see fig.~\ref{fig:millenniumhier}). 
The recognition of the {\it Cosmic Web} as a key aspect in the emergence of structure in the Universe came with 
early analytical studies and approximations concerning the emergence of structure out of a nearly featureless 
primordial Universe. In this respect the Zel'dovich formalism~\citep{zeldovich1970} played a seminal role. It 
led to the view of structure formation in which planar pancakes form first, draining into filaments which in turn 
drain into clusters, with the entirety forming a cellular network of sheets. 

The Megaparsec scale tidal shear forces are the main agent for the contraction of matter into the sheets and 
filaments which trace out the cosmic web. The anisotropic contraction of patches of matter depends sensitively 
on the signature of the tidal shear tensor eigenvalues. With two positive eigenvalues and one negative,  
$(-++)$, we will see strong collapse along two directions. Dependent on the overall overdensity, along the 
third axis collapse will be slow or not take place at all. Likewise, a sheetlike membrane will be the product of 
a $(--+)$ signature, while a $(+++)$ signature inescapably leads to the full collapse of a density peak into 
a dense cluster. 

For a proper understanding of the Cosmic Web we need to invoke two important observations stemming from intrinsic  
correlations in the primordial stochastic cosmic density field. When restricting ourselves to overdense regions in 
a Gaussian density field we find that mildly overdense regions do mostly correspond to filamentary $(-++)$ tidal 
signatures \citep{pogosyan1998}. This explains the prominence of filamentary structures in the cosmic Megaparsec 
matter distribution, as opposed to a more sheetlike appearance predicted by the Zeld'ovich theory. 
The same considerations lead to the finding that the highest density regions are mainly confined to density peaks 
and their immediate surroundings. 

The second, most crucial, observation \citep{bondweb96} is the intrinsic link between filaments and 
cluster peaks. Compact highly dense massive cluster peaks are the main source of the Megaparsec tidal force 
field: filaments should be seen as tidal bridges between cluster peaks. This may be directly understood by realizing 
that a $(-++)$ tidal shear configuration implies a quadrupolar density distribution (eqn.~\ref{eq:quadtide}). This 
means that an evolving filament tends to be accompanied by two massive cluster patches at its tip. These overdense 
protoclusters are the source of the specified shear, explaining the canonical {\it cluster-filament-cluster} configuration 
so prominently recognizable in the observed Cosmic Web. 
 
\section{the Cosmic Web and Galaxy Rotation: MMF analysis}
With the cosmic web as a direct manifestation of the large scale tidal field we may wonder whether we can 
detect a connection with the angular momentum of galaxies or galaxy halos. In section~\ref{sec:tidaltorque} we 
have discussed how tidal torques generate the rotation of galaxies. Given the common tidal origin we would 
expect a significant correlation between the angular momentum of halos and the filaments or sheets in which 
they are embedded. It was \cite{LeePen00} who pointed out that this link should be visible in alignment of the 
spin axis of the halos with the inducing tidal tensor, and by implication the large scale environment in which 
they lie. 

In order to investigate this relationship it is necessary to isolate filamentary features in the cosmic 
matter distribution. A systematic morphological analysis of the cosmic web has proven to be a far from trivial problem, though there have recently been some significant advances.  Perhaps the most rigorous program, with a particular emphasis on the 
description and analysis of filaments, is that of the {\it skeleton} analysis of density fields by 
\cite{novikov2006,sousbie2007}.  Another strategy has been followed by \cite{hahn2007} who identify clusters, filaments, 
walls and voids in the matter distribution on the basis of the tidal field tensor $\partial^2 \phi/\partial x_i \partial x_j$, 
determined from the density distribution filtered on a scale of $\approx 5h^{-1}\hbox{\rm Mpc}$. 
\begin{figure*}[t]
	\includegraphics[width=12.0cm]{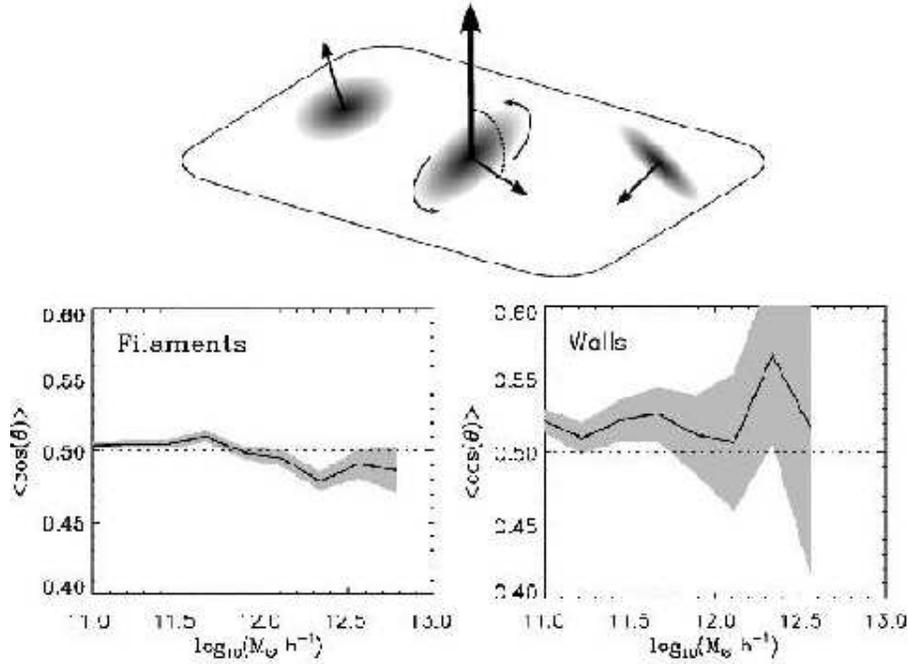}
	\caption{Average alignment angle $\cos\theta$ between the halo spin direction and the orientation of the host structure as a function of halo mass, for filaments (left) and walls (right) in a $\Lambda$CDM N-body simulation. Filaments and walls were identified using the MMF technique. The dotted line indicates a uniform distribution of halo orientations.  The shaded area corresponds to the standard deviation of 1000 random realisations with the same number of galaxies as the halo sample and is wider in the case of walls due to the lower number of haloes in walls. From \citet{aragon2007}.} 
\label{fig:wallspin}
\end{figure*}

The one method that explicitly takes into account the hierarchical nature of the mass distribution when analyzing the weblike 
geometries is the Multiscale Morphology Filter (MMF), introduced by \cite{aragonmmf2007}. The MMF dissects the cosmic web on 
the basis of the multiscale analysis of the Hessian of the density field. It starts by translating an N-body particle 
distribution or a spatial galaxy distribution into a DTFE density field \citep[see][]{weyschaap2007}. 
This guarantees a morphologically unbiased and optimized density field retaining all features 
visible in a discrete galaxy or particle distribution. The DTFE field is filtered over a range of 
scales. By means of morphology filter operations defined on the basis of the Hessian of the filtered density fields 
the MMF successively selects the regions which have a bloblike (cluster) morphology, a filamentary morphology and a 
planar morphology, at the scale at which the morphological signal is optimal. By means of a percolation criterion the 
physically significant filaments are selected. Following a sequence of blob, filament and wall 
filtering finally produces a map of the different morphological features in the particle 
distribution. 

With the help of the MMF we have managed to find the relationship of shape (inertia tensor) and spin-axis 
of halos in filaments and walls and their environment. On average, the long axis of filament halos 
is directed along the axis of the filament; wall halos tend to have their longest axis in the 
plane of the wall. At the present cosmic epoch the effect is stronger for massive halos. Interestingly, 
the trend appears to change in time: low mass halos tended to be more strongly aligned but as time proceeds local 
nonlinear interactions affect the low mass halos to such an extent that the situation has reversed.  

The orientation of the rotation axis provides a more puzzling picture (fig.~\ref{fig:wallspin}). The rotation axis of 
low mass halos tends to be directed along the filament's axis while that of massive halos appears to align in the 
perpendicular direction. In walls there does not seem to exist such a bias: the rotation-axis of both 
massive and light haloes tends to lie in the plane of the wall. At earlier cosmic epochs the trend 
in filaments was entirely different: low mass halo spins were more strongly aligned as large scale 
tidal fields were more effective in directing them. During the subsequent evolution in high-density 
areas, marked by strongly local nonlinear interactions with neighbouring galaxies, the alignment 
of the low mass objects weakens and ultimately disappears. 

\section{Tidal Fields and Void alignment}
A major manifestation of large scale tidal influences is that of the
alignment of shape and angular momentum of objects \citep[see][]{bondweb96, 
desjacques2007}.  The alignment of the orientations of galaxy haloes, 
galaxy spins and clusters with larger scale structures such as clusters, 
filaments and superclusters has been the subject of numerous studies 
\citep[see e.g.][]{binggeli1982, bond1987, rhee1991, plionis2002, basilakos2006, 
trujillo2006, aragon2007,leevrard2007,leespringel2007}. 

Voids are a dominant component of the Cosmic Web \citep[see
e.g.][]{tully2007, weyrom2007}, occupying most of the volume of space.
Recent analytical and numerical work \citep{parklee2007,leepark2007,platen2008} 
discussed the magnitude of the tidal contribution to the shape and alignment of 
voids. \citet{leepark2007} found that the ellipticity distribution of voids is a 
sensitive function of various cosmological parameters and remarked that the shape 
evolution of voids provides a remarkably robust constraint on the dark energy equation 
of state. \cite{platen2008} presented evidence for significant alignments between
neigbouring voids, and established the intimate dynamic link between
voids and the cosmic tidal force field. 

Voids were identified with the help of the Watershed Void Finder (WVF) 
procedure \citep{platen2007}.  The WVF technique is based on
the topological characteristics of the spatial density field and
thereby provides objectively defined measures for the size, shape and
orientation of void patches.

\subsection{Void-Tidal Field alignments: formalism}
In order to trace the contributions of the various scales to the void
correlations \cite{platen2008} investigated the alignment between the void shape 
and the tidal field smoothed over a range of scales $R$.  The alignment function 
${\mathcal A}_{TS}(R_1)$ between the local tidal field tensor $T_{ij}(R_1)$, Gaussian 
filtered on a scale $R_1$ at the void centers, and the void shape ellipsoid is determined as 
follows. 
\begin{figure*}
	\mbox{\hskip -0.1truecm\includegraphics[width=12.0cm]{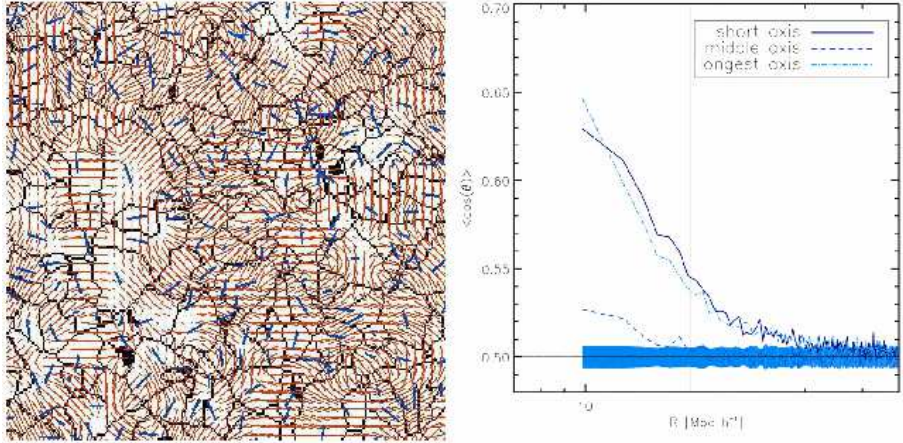}}
\caption{Left: on the landscape with WVF void boundaries the tidal field compressional component is
  represented by tidal bars (red), representing the direction and strength of the tidal field. Also depicted are the void shape bars
  (blue). Right: the dotted line shows $\mathcal{C}_{TS}$, the alignment
     between the compressional direction of the tidal field and the
     shortest shape axis. For comparison the short axis alignment is also superimposed. 
     From Platen et al. 2008.}
     \label{fig:cor2}
\end{figure*}
For each individual void region the shape-tensor $\mathcal{S}_{ij}$ is calculated by 
summing over the $N$ volume elements $k$ located within the void,
\begin{eqnarray}
  \qquad \mathcal{S}_{ij}& = - &\sum_{k} x_{ki} x_{kj} \qquad \ \ \ \ \ \ \ \ \textrm{(offdiagonal)} \\
  \qquad \mathcal{S}_{ii}& =   &\sum_{k} \left({\bf x}_k^2-x_{ki}^2\right) \qquad\ \ \textrm{(diagonal)}\,,\nonumber 
\end{eqnarray}
where ${\bf x}_k$ is the position of the $k$-th volume element within the void, with respect to the (volume-weighted)  
void center ${\bf {\overline{r}}}_v$, i.e. ${\bf x}_k = {\bf r}_k - {\bf {\overline{r}}}_v$.  The shape tensor $\mathcal{S}_{ij}$ is 
related to the inertia tensor $\mathcal{I}_{ij}$.  However, it differs in assigning equal weight to each volume element within the 
void region. Instead of biasing the measure towards the mass concentrations near the edge of voids, the shape tensor $\mathcal{S}_{ij}$ 
yields a truer reflection of the void's interior shape. 

The smoothing of the tidal field is done in Fourier space using a Gaussian window function ${\hat W}^*({\bf k};R)$:
\begin{eqnarray}
T_{ij}({\bf r};R)\,=\,
{\displaystyle 3 \over \displaystyle 2}
\Omega H^2 \int \d3k\left({\displaystyle k_i k_j \over \displaystyle k^2}-{\displaystyle 1 \over \displaystyle 3} 
\delta_{ij}\right)\,{\hat W}^*({\bf k};R)\,{\hat \delta}({\bf k})\,
{\rm e}^{-{\rm i}{\bf k}\cdot{\bf r}}&&\, \nonumber
\end{eqnarray} 
Here, ${\hat \delta}({\bf k})$ is the Fourier amplitude of the relative density fluctuation field at wavenember $\bf k$.

Given the void shape ${\mathcal S}_{ij}$ and the tidal tensor $T_{ij}$, for every void the function ${\Gamma}_{TS}(m,R_1)$ 
at the void centers is determined:
\begin{eqnarray}
{\Gamma}_{TS}(m;R_1)&\,=\,&\ -\ {\displaystyle \sum\nolimits_{i,j} 
{\tilde {\mathcal S}}_{m,ij}\,T_{ij}({\bf r}_m;R_1) 
\over \displaystyle {\tilde {\mathcal S}}_m\,\,T({\bf r}_m;R_1)}
\end{eqnarray}
where $T({\bf r}_m;R_1)$ is the norm of the tidal tensor $T_{ij}({\bf r_m})$ filtered on a scale $R_1$ and 

The void-tidal alignment ${\mathcal A}_{TS}(R_1)$ at a scale $R$ is then the ensemble average 
\begin{eqnarray}
{\mathcal A}_{TS}(R_1)&\,=\,&\langle\,\Gamma_{TS}(R_1)\,\rangle\,.
\label{eq:ats}
\end{eqnarray}
which we determine simply by averaging $\Gamma_{TS}(m,R_1)$ over the complete sample of voids. 

\subsection{Void-Tidal Field alignments: results}
A visual impression of the strong relation between the void's shape
and orientation and the tidal field is presented in the lefthand panel 
of fig~\ref{fig:cor2} (from \cite{platen2008}). The tidal field configuration is 
depicted by means of (red-coloured) tidal bars. These bars represent the compressional 
component of the tidal force field in the slice plane, and have a size proportional to 
its strength and are directed along the corresponding tidal axis. The bars are superimposed
on the pattern of black solid watershed void boundaries, whose
orientation is emphasized by means of a bar directed along the
projection of their main axis.
 
The compressional tidal forces tend to be directed perpendicular to
the main axis of the void. This is most clearly in regions where the
forces are strongest and most coherent. In the vicinity of great
clusters the voids point towards these mass concentrations, stretched
by the cluster tides. The voids that line up along filamentary
structures, marked by coherent tidal forces along their ridge, are
mostly oriented along the filament axis and perpendicular to the local
tidal compression in these region.  The alignment of small voids along
the diagonal running from the upper left to the bottom right is
particularly striking.
 
A direct quantitative impression of the alignment between the void shape and tidal field, may be obtained 
from the righthand panel of fig.~\ref{fig:cor2}. The figure shows $\mathcal{C}_{TS}$ (dotted line), the alignment 
between the compressional direction of the tidal field and the shortest shape axis. It indicates that the tidal field 
is instrumental in aligning the voids. To further quantify and trace the tidal origin of the alignment one can 
investigate the local shape-tide alignment function ${\mathcal A}_{TS}$ (eqn.~\ref{eq:ats}) versus 
the smoothing radius $R_1$.

This analysis reveals that the alignment remains strong over the whole range of
smoothing radii out to $R_1 \approx 20-30h^{-1}\hbox{\rm Mpc}$ and peaks at 
a scale of $R_1 \approx 6h^{-1} \hbox{\rm Mpc}$. This scale is very close to the average 
void size, and also close to the scale of nonlinearity. This is not a coincidence: the identifiable voids 
probe the linear-nonlinear transition scale. The remarkably strong alignment signal at large radii than 
$R_1>20h^{-1}\hbox{\rm Mpc}$ (where ${\mathcal A}_{TS}\approx 0.3$), can only be understood if 
large scale tidal forces play a substantial role in aligning the voids. 
 
\section{Final remarks}
The last word on the origin of galactic angular momentum has not been said yet.  It is now a part of our cosmological paradigm that the global tidal fields from the irregular matter distribution on all scales is the driving force, but the details of how this works have yet to be explored.  That is neither particularly demanding nor particularly difficult, it is simply not trendy: there are other problems of more pressing interest.  The transition from shear dominated to rotation dominated motion is hardly explored and will undoubtedly be one of the principal by-products of cosmological simulations with gas dynamics and star formation.
 
The role of tidal fields has been found to be more profound than the mere transfer of angular momentum to proto-objects.  The cosmic tidal fields evidently shape the entire distribution and dynamics of galaxies: they shape what has become known as the ``cosmic web''. Although we see angular momentum generation in cosmological N-Body simulations it is not clear that the simulations do much more than tell us what happened: galaxy haloes in N-Body models have acquired spin by virtue of tidal interaction.  We draw comfort from the fact that the models give the desired result.  

Nikos Voglis' approach was somewhat deeper: he wanted to understand things at a mechanistic level rather than simply to simulate them and observe the result.  In that he stands in the finest tradition of the last of the great Hellenistic scientists, Hipparchus of Nicaea, who studied motion of bodies under gravity.  Perhaps we should continue in the spirit of Nikos' work by trying to understand things rather than simply simulate them.
 
Nikos was a good friend, a fine scientist and certainly one of the kindest people one could ever meet. It was less than one 
year ago when we met for the last time at the {\it bernard60} conference in Valencia.  We were of course delighted to see him and we shall cherish that brief time together.
 
\section{Acknowledgments}
We thank Panos Patsis and George Contopoulos for the opportunity of delivering this tribute to our late friend. We are 
grateful to Volker Springel for allowing us to use figure 3. We particularly 
wish to acknowledge Miguel Arag\'on-Calvo and Erwin Platen for allowing us to use their scientific results:  
their contributions and discussions have been essential for our understanding of the subject.


\begin{thebibliography}{99}
\bibitem[\protect\citeauthoryear{Arag\'on-Calvo et al.}{2007}]{aragon2007} 
   {Arag\'on-Calvo} M.A., {Jones} B.J.T., {van de Weygaert} R., {van der Hulst} J.M.:
   Astrophys. J. \textbf{655}, L5 (2007)
\bibitem[\protect\citeauthoryear{Arag\'on-Calvo et al.}{2007}]{aragonmmf2007} 
  {Arag\'on-Calvo} M.A., {Jones} B.J.T., {van de Weygaert} R., {van der Hulst} J.M.: 
  Astron. Astrophys. \textbf{474}, 315 (2007)
\bibitem[\protect\citeauthoryear{Barnes \& Efstathiou}{1987}]{Barnes87} 
  {Barnes} J., {Efstathiou} G.: Astrophys. J. \textbf{319}, 575 (1987)
\bibitem[\protect\citeauthoryear{Basilakos et al.}{2006}]{basilakos2006}
  {Basilakos} S., {Plionis} M., {Yepes} G., {Gottl\"ober} S., {Turchaninov} V.: 
  Mon. Not. Roy. Astron. Soc. \textbf{365}, 539 (2006)
\bibitem[\protect\citeauthoryear{Bett et al.} {2007}]{Bett07} 
  {Bett} P., {Eke} V., {Frenk} C.S., {Jenkins} A., {Helly} J., {Navarro} J.: 
  Mon. Not. Roy. Astron. Soc. \textbf{376}, 215 (2007)
\bibitem[\protect\citeauthoryear{Binggeli}{1982}]{binggeli1982}
  {Binggeli} B.: Astron. Astrophys. \textbf{107}, 338 (1982)
\bibitem[\protect\citeauthoryear{Bond}{1987}]{bond1987}
  {Bond} J.R.: in \textit{Nearly Normal Galaxies}, 
  ed. S. Faber (Springer, New York), p. 388 (1987)
\bibitem[\protect\citeauthoryear{Bond et al.}{1996}]{bondweb96} 
  {Bond} J.~R., {Kofman} L., {Pogosyan} D.: Nature \textbf{380}, 603 (1996)
\bibitem[\protect\citeauthoryear{Catelan \& Theuns}{1999}]{CatTheuns99}
  {Catelan}, P., and {Theuns}, T.: Mon. Not. Roy. Astron. Soc. \textbf{282}, 436 (1999)
\bibitem[\protect\citeauthoryear{Desjacques}{2007}]{desjacques2007} 
  {Desjacques} V.: arXiv0707.4670 (2007)
\bibitem[\protect\citeauthoryear{Doroshkevich}{1970}]{dorosh1970} 
  {Doroshkevich} A.G.: Astrophysics \textbf{6}, 320 (1970)
\bibitem[\protect\citeauthoryear{Efstathiou \& Jones}{1979}]{efstathjones79} 
  {Efstathiou} G., {Jones} B.~J.~T.: Mon. Not. Roy. Astron. Soc. \textbf{186}, 133 (1979)
\bibitem[\protect\citeauthoryear{Feynman}{1970}]{Feynman70} 
	{Feynman}, R.P.: \textit{Feynman Lectures On Physics}, (Addison Wesley Longman) (1970)
\bibitem[\protect\citeauthoryear{Hahn \etal}{2007}]{hahn2007}
  {Hahn} O., {Porciani} C., {Carollo} M., {Dekel} A.: Mon. Not. R. Astron. Soc. 
  \textbf{375}, 489 (2007) 
\bibitem[\protect\citeauthoryear{Harrison} {1971}]{harrison1971}
	{Harrison}, E.R.: Mon. Not. Roy. Astron. Soc. \textbf{154}, 167 (1971)
\bibitem[\protect\citeauthoryear{Hirata \& Seljak}{2004}]{HirSel04}
	{Hirata} C.M., {Seljak} U.: Phys. Rev. D, 70, 063526 (2004)
\bibitem[\protect\citeauthoryear{Hoyle}{1949}]{Hoyle49}
	{Hoyle} F.:  in \textit{Problems of Cosmical Aerodynamics}, ed. J. M. Burgers
	and H. C. van de Hulst (Dayton: Central Air Documents Office), 195 (1949)
\bibitem[\protect\citeauthoryear{Icke}{1973}]{icke1973}
  {Icke} V.: Astron. Astrophys. \textbf{27}, 1 (1973)
\bibitem[\protect\citeauthoryear{Jones}{1976}]{Jones76} 
	{Jones} B.J.T.: Rev. Mod. Phys. \textbf{48}, 107 (1976)
\bibitem[\protect\citeauthoryear{Jones \& Efstathiou} {1979}]{jonesefstath79} 
   {Jones} B.~J.~T., {Efstathiou} G.: Mon. Not. Roy. Astron. Soc. \textbf{189}, 27 (1979)
\bibitem[\protect\citeauthoryear{Lee \& Evrard}{2007}]{leevrard2007}
  {Lee} J., {Evrard} A.E.: Astrophys. J. \textbf{657}, 30 (2007)
\bibitem[\protect\citeauthoryear{Lee \& Park}{2007}]{leepark2007}
  {Lee} J., {Park} D.: arXiv0704.0881 (2007)
\bibitem[\protect\citeauthoryear{Lee \& Pen}{2000}]{LeePen00}
	{Lee} J., {Pen} U.: Astrophys. J. \textbf{532}, L5 (2000) 
\bibitem[\protect\citeauthoryear{Lee \& Pen}{2001}]{LeePen01}
	{Lee} J., {Pen} U.: Astrophys. J. \textbf{555}, 106 (2001) 
\bibitem[\protect\citeauthoryear{Lee \& Pen}{2007}]{LeePen07}
	{Lee} J., {Pen} U.: arXiv0707.1690 (2007)
\bibitem[\protect\citeauthoryear{Lee et al.}{2007}]{leespringel2007}
  {Lee}~J., {Springel}~V., {Pen}~U-L., {Lemson}~G.: arXiv0709.1106 (2007)
\bibitem[\protect\citeauthoryear{Massey et al.}{2007}]{massey2007} 
  {Massey} R., {Rhodes} J., {Ellis} R., {Scoville} N., 
  {Leauthaud} A., {Finoguenov} A., {Capak} P., {Bacon} D., {Aussel} H., 
  {Kneib} J.-P., {Koekemoer} A., {McCracken} H., {Mobasher} B., {Pires} S., 
  {Refregier} A., {Sasaki} S., {Starck} J.-L., {Taniguchi} Y., 
  {Taylor} A., {Taylor} J.: Nature \textbf{445}, 286 (2007)
\bibitem[\protect\citeauthoryear{Novikov, Colombi \& Dor\'e}{2006}]{novikov2006} 
  {Novikov} D., {Colombi} S., {Dor\'e} O.: Mon. Not. R. Astron. Soc. \textbf{366}, 1201 (2006)
\bibitem[\protect\citeauthoryear{Oort}{1970}]{oort1970} 
	{Oort} J.H.: Astron. Astrophys. \textbf{7}, 381 (1970)
\bibitem[\protect\citeauthoryear{Palmer \& Voglis}{1983}]{PalVog83}
	{Palmer} P.L., {Voglis} N.: Mon. Not. Roy. Astron. Soc. \textbf{205}, 543 (1983)
\bibitem[\protect\citeauthoryear{Park \& Lee}{2007}]{parklee2007}
  {Park} D. and {Lee} J.: Astrophys. J. \textbf{665}, 96 (2007)
\bibitem[\protect\citeauthoryear{Peebles}{1969}]{Peebles69} 
	{Peebles} P. J. E.: Astrophys. J. \textbf{155}, 393 (1969)
\bibitem[\protect\citeauthoryear{Peebles}{1980}]{peebles1980}
  {Peebles} P.~J.~E.: \textit{The large-scale structure of the
  universe}, (Princeton University Press) (1980)
\bibitem[\protect\citeauthoryear{Platen, van de Weygaert \& Jones}{2007}]{platen2007}
  {Platen} E., {van de Weygaert} R., {Jones} B.J.T.: Mon. Not. Roy. Astron. Soc. \textbf{380}, 551 (2007)
\bibitem[\protect\citeauthoryear{Platen, van de Weygaert \& Jones}{2008}]{platen2008} 
  {Platen} E., {van de Weygaert} R., {Jones} B.J.T.: 
  Mon. Not. R. Astron. Soc., in press (2008)
\bibitem[\protect\citeauthoryear{Plionis \& Basilakos}{2002}]{plionis2002}
  {Plionis} M., {Basilakos} S.: Mon. Not. Roy. Astron. Soc. \textbf{329}, L47 (2002)
\bibitem[\protect\citeauthoryear{Pogosyan et al.}{1998}]{pogosyan1998} 
  {Pogosyan} D.Yu.m {Bond} J.R., {Kofman} L., {Wadsley} J.: Cosmic Web: 
  Origin and Observables. In: \textit{Wide Field Surveys in 
  Cosmology, 14th IAP meeting), eds. S. Colombi, Y. Mellier, (Editions Frontieres}, p. 61 (1998)
\bibitem[\protect\citeauthoryear{Porciani et al.}{2002}]{Porciani02} 
   {Porciani} C., {Dekel} A., {Hoffman} Y.: Mon. Not. Roy. Astron. Soc. \textbf{332}, 325 (2002)
\bibitem[\protect\citeauthoryear{Rhee et al.}{1991}]{rhee1991}
  {Rhee} G., {van Haarlem} M., {Katgert} P.: Astron. J. \textbf{103}, 6 (1991)
\bibitem[\protect\citeauthoryear{Romano-D\'{\i}az \& van de Weygaert}{2007}]{weyrom2007} 
  {Romano-D\'{\i}az} E., {van de Weygaert} R.: Mon. Not. Roy. Astron. Soc. \textbf{382}, 2 (2007)
\bibitem[\protect\citeauthoryear{Sousbie et al.}{2007}]{sousbie2007} 
   {Sousbie} T., {Pichon} C., {Colombi} S., {Novikov} D., {Pogosyan} D.: astroph/07073123 (2007)
\bibitem[\protect\citeauthoryear{Springel et al.}{2005}]{springmillen2005} 
  {Springel} V., {White} S.D.M., {Jenkins} A., {Frenk} C.S., {Yoshida} N., {Gao} L., 
  {Navarro} J., {Thacker} R., {Croton} D., {Helly} J., {Peacock} J.A., {Cole} S., 
  {Thomas} P., {Couchman} H., {Evrard} A., {Colberg} J.M., {Pearce} F.: Nature \textbf{435}, 
  629 (2005)
\bibitem[\protect\citeauthoryear{Trujillo et al.}{2006}]{trujillo2006} 
  {Trujillo} I., {Carretero} C., {Patiri} S.~G.: Astrophys. J. \textbf{640}, L111 (2006)
\bibitem[\protect\citeauthoryear{Tully et al.}{2007}]{tully2007}
  {Tully} R.B., {Shaya} E.J., {Karachentsev} I.D., {Courtois} H., {Kocevski} D.D., 
  {Rizzi} L., {Peel} A.: arXiv:0705.4139 (2007)
\bibitem[\protect\citeauthoryear{van den Bosch et al.}{2002}]{Bosch02} 
   {van den Bosch} F.~C., {Abel} T., {Croft} R.~A.~C., {Hernquist} L., {White} S.~D.~M.: 
   Astrophys. J. \textbf{576}, 21 (2002)
\bibitem[\protect\citeauthoryear{van de Weygaert \& Bertschinger}{1996}]{weyedb1996} 
  {van de Weygaert} R., {Bertschinger} E.: Mon. Not. R. Astron. Soc. \textbf{281}, 84 (1996)
\bibitem[\protect\citeauthoryear{van de Weygaert \& Schaap}{2007}]{weyschaap2007} 
  {van de Weygaert} R., {Schaap} W.E: The Cosmic Web: Geometric Analysis. In:
  \textit{Data Analysis in Cosmology}, lectures summerschool Valencia 2004, 
  ed. by V. Mart{\'{\i}}nez, E. Saar, E. Mart{\'{\i}}nez-Gonz\'alez, M. Pons-Borderia 
  (Springer-Verlag), 129pp. (2008)
\bibitem[\protect\citeauthoryear{van de Weygaert \& Bond}{2008}]{weybondgh2008} 
  {van de Weygaert} R., {Bond} J.R.: Clusters and the Theory of the Cosmic Web. In: 
  \textit{A Pan-Chromatic View of Clusters of Galaxies and the LSS}, 
  eds. M. Plionis, D. Hughes, O. L\'opez-Cruz (Springer 2008)
\bibitem[\protect\citeauthoryear{White}{1984}]{White84}	
	{White} S. D. M.: Astrophys. J. \textbf{286}, 38 (1984)
\bibitem[\protect\citeauthoryear{Zeldovich}{1970}]{zeldovich1970} 
  {Zel'dovich} Ya. B.: Astron. Astrophys. \textbf{5}, 84 (1970)
\end{thebibliography}
\end{document}